\documentclass[traditabstract,rnote]{aa}  
\usepackage{graphicx}
\usepackage{txfonts}
\begin{document}
\title{A conjugate gradient method for the solution of the non--LTE
  line radiation transfer problem}

   \author{F. Paletou
          \and
          E. Anterrieu
          }

          \institute{Laboratoire d'Astrophysique de Toulouse--Tarbes, Universit\'e de Toulouse, CNRS, 14 av. E. Belin, 31400 Toulouse, France\\
            \email{fpaletou@ast.obs-mip.fr} }

   \date{Received May 14, 2009; accepted August 26, 2009}

 
  \abstract
{  This study concerns the fast and accurate solution of the line
    radiation transfer problem, under non-LTE conditions.
  We propose and evaluate an alternative iterative scheme to the
    classical ALI-Jacobi method, and to the more recently proposed
    Gauss-Seidel and Successive Over-Relaxation (GS/SOR) schemes.
  Our study is indeed based on the application of a preconditioned
    bi-conjugate gradient method (BiCG-P).
  Standard tests, in 1D plane parallel geometry and in the frame of
    the two-level atom model, with monochromatic scattering, are
    discussed. Rates of convergence between the previously mentioned
    iterative schemes are compared, as well as their respective timing
    properties. The smoothing capability of the BiCG-P method is also
    demonstrated. }

   \keywords{Radiative transfer -- Methods: numerical
               }

   \maketitle
%

\section{Introduction}

The solution of the radiative transfer equation, under non--LTE
conditions, is a classical problem in astrophysics. Many numerical
methods have been used since the beginning of the computer era, from
difference equations methods (e.g., Feautrier 1964, Cuny 1967, Auer \&
Mihalas 1969) to iterative schemes (e.g., Cannon 1973, Scharmer 1981,
Olson et al. 1986).

Since the seminal paper of Olson et al. (1986), the so-called ALI or
\emph{Accelerated $\Lambda$-Iteration} scheme is nowadays one of the
most popular method for solving complex radiation transfer
problems. Using as approximate operator the diagonal of the $\Lambda$
operator (see e.g., Mihalas 1978), it is a \emph{Jacobi} iterative
scheme. It has been generalized for multilevel atom problems (Rybicki
\& Hummer 1991), multi-dimensional geometries (Auer \& Paletou 1994,
Auer et al. 1994, van Noort et al. 2002), and polarized radiation
transfer (e.g., Faurobert et al. 1997, Trujillo Bueno \& Manso Sainz
1999).

Despite the many success of the ALI method, Trujillo Bueno \& Fabiani
Bendicho (1995) proposed \emph{a novel iterative scheme} for the
solution of non--LTE radiation transfer. They adapted Gauss-Seidel and
Successive Over-Relaxation (GS/SOR) iterative schemes, well-known in
applied mathematics as being \emph{superior}, in terms of convergence
rate, to the ALI--Jacobi iterative scheme. It is interesting to note
that, besides their application to radiation transfer, both Jacobi and
GS/SOR iterative schemes were proposed during the {\sc xix}$^{th}$
century.

Conjugate gradient-like, hereafter CG-like, iterative methods were
proposed more recently by Hestenes \& Stiefel (1952). They are of a
distinct nature than ALI-Jacobi and GS/SOR methods. Unlike the later,
CG-like methods are so-called \emph{non-stationary} iterative methods
(see e.g., Saad 2008). Very few articles discussed its application to
the radiation transfer equation (see e.g., Amosov \& Dmitriev 2005)
and, to the best of our knowledge, they have not been properly
evaluated, so far, for astrophysical purposes.

Such an evaluation is the scope of the present research note. This
alternative approach is relevant to the quest for ever faster and
accurate numerical methods for the solution of the non--LTE line
radiation transfer problem.


\section{The iterative scheme}

In the \emph{two-level atom case}, and with complete frequency
redistribution, the non-LTE line source function is usually written as

\begin{equation}
S(\tau) = (1 - \varepsilon) \bar{J}(\tau) + \varepsilon B(\tau) \, ,
\label{eq:s2n}
\end{equation}
where $\tau$ is the optical depth, $\varepsilon$ is the collisional
destruction probability\footnote{The albedo $a=(1-\varepsilon)$ is more
  commonly used in general studies of the radiation transfer
  equation.}, $B$ is the Planck function and $\bar{J}$ is the usual
mean intensity:

\begin{equation}
\bar{J} = \oint \displaystyle{{{d \Omega} \over {4 \pi}}}
  \int_{-\infty}^{+\infty} {\phi_{\nu} I_{\nu \Omega} d \nu} \, ,
\label{eq:jbar2}
\end{equation}
where the optical depth dependence of the specific intensity $I_{\nu
  \Omega}$ (and, possibly, of the line absorption profile
$\phi_{\nu}$) is omitted for simplicity.

The mean intensity is usually written as the \emph{formal} solution of the
radiative transfer equation

\begin{equation}
\bar{J} = \Lambda [S] \, ,
\end{equation}
or, in other terms, the solution of the radiation transfer equation
for a \emph{known} source function (see e.g., Mihalas 1978).

Let us define $A \equiv {\rm Id} - (1-\varepsilon)\Lambda$, the unknown $x
\equiv S$ and the right-hand side $b \equiv \varepsilon B$. Then, the
solution of the radiative transfer equation is equivalent to solving
the system of equations:

\begin{equation}
Ax = b \, .
\end{equation}

\begin{figure}
  \centering
  \includegraphics[width=10cm,angle=0]{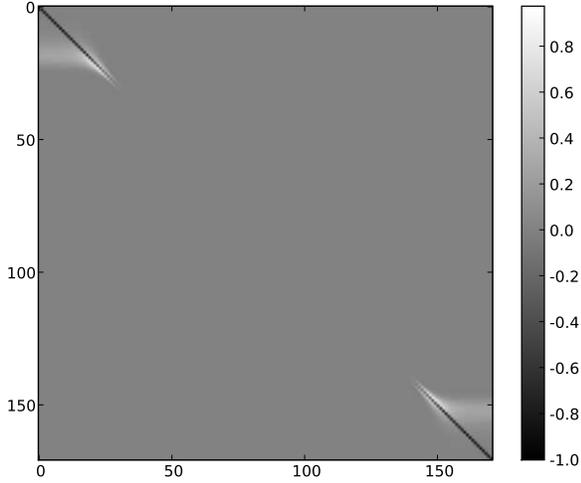}
  \caption{Typical structure of the $A$ operator, computed for the
    case of a self-emitting 1D, plane parallel finite slab. Such an
    operator is, more generally, non-symmetric positive definite.
      For the sake of getting a better contrast for the off-diagonal
      parts, we magnified the negative values of the original operator
      by a factor of 5, and we display in fact the opposite of the
      modified operator.}
  \label{Fig1}
\end{figure}

Among the various forms of CG-based methods, the bi-conjugate gradient
(hereafter BiCG) method is more appropriate for the case of operators
$A$ which are not symmetric positive definite. It is indeed the case
for our radiative transfer problem, as illustrated in Fig. 1 where we
display the structure of the operator $A$ for the case of a symmetric
1D grid of maximum optical depth $\tau_{\rm max}=2 \times 10^{8}$,
discretized with 8 spatial points per decade and using
$\varepsilon=10^{-6}$.

A general description of the BiCG gradient method can be found in
several classical textbooks of applied mathematics (see e.g., Saad
2008; it is important to note that this method requires the use of the
transpose of the $A$ operator, $A^{T}$). However, the BiCG method
remains efficient until the $A$ operator becomes very badly
conditioned i.e., for very high albedo cases, typically when
$\varepsilon < 10^{-6}$. For the later cases, it is therefore crucial
to switch to a more efficient scheme, using preconditioning of the
system of equations.

Hereafter, we derive the algorithm for the BiCG method with
preconditioning (hereafter BiCG-P). In such a case, one seeks instead
for a solution of:

\begin{equation}
M^{-1} A x = M^{-1} b \, ,
\end{equation}
where the matrix $M$ should be easy to invert. A ``natural'' choice
for the non-LTE radiation transfer problem, is to precondition the
system of equations with the \emph{diagonal} of the full operator $A$.

From an initial guess for the source function, $x_{0}=S^{(0)}$, compute a
\emph{residual} $r_{0}$ such as:

\begin{equation}
r_{0} = b - Ax_{0} \, .
\end{equation}
Set the second residual $\tilde{r}_{0}$ such that
$(r_{0},\tilde{r}_{0}) \neq 0$, where $(a,b)$ means the \emph{inner
  product} between vectors $a$ and $b$. For all the cases considered by
us, setting $r_{0}=\tilde{r}_{0}$ proved to be adequate.

Now, the BiCG-P algorithm consists in running until convergence the
following iterative scheme, where $j$ is the iterative scheme
index. The first steps consist in solving:

\begin{equation}
M z_{j-1} = r_{j-1} \, ,
\label{eq:start}
\end{equation}
and

\begin{equation}
M^{T} \tilde{z}_{j-1} = \tilde{r}_{j-1} \, .
\end{equation}
These steps are both simple and fast to compute, since $M$ is a
diagonal operator.
Then one computes:

\begin{equation}
\rho_{j-1} = (z^{T}_{j-1},\tilde{r}_{j-1}) \, .
\end{equation}
If $\rho_{j-1}=0$ then the method fails, otherwise the algorithm
continues with the following computations. If $j=1$ then, set
$p_{j}=z_{j-1}$ and $\tilde{p}_{j}=\tilde{z}_{j-1}$. For $j>1$,
compute:
 
\begin{equation}
\beta_{j-1} = \rho_{j-1}/\rho_{j-2} \, ,
\end{equation}
 
\begin{equation}
p_{j} = z_{j-1} + \beta_{j-1} p_{j-1} \, ,
\end{equation}
and:
 
\begin{equation}
\tilde{p}_{j} = \tilde{z}_{j-1} + \beta_{j-1} \tilde{p}_{j-1} \, .
\end{equation}
Then make $q_{j}=Ap_{j}$ and  $\tilde{q}_{j}=A^{T} \tilde{p}_{j}$, as well as:

\begin{equation}
\alpha_{j} = { {\rho_{j-1}} \over {({\tilde{p}^{T}_{j}},{q_{j}})} } \, .
\end{equation}
Finally, one advances the source function according to:

\begin{equation}
x_{j} = x_{j-1} + \alpha_{j} p_{j} \, ,
\end{equation}
and the two residuals:

\begin{equation}
r_{j} = r_{j-1} - \alpha_{j} q_{j} \, ,
\end{equation}
and,

\begin{equation}
\tilde{r}_{j} = \tilde{r}_{j-1} - \alpha_{j} \tilde{q}_{j} \, .
\label{eq:end}
\end{equation}

This process, from Eq. (\ref{eq:start}) to Eq. (\ref{eq:end}), is then
repeated to convergence. We have found that a convenient stopping criterion
is to interrupt the iterative process when:

\begin{equation}
{ {\| r \|_{2}} \over {\| b \|_{2}}} < 10^{-2} \, .
\label{eq:stop_crit}
\end{equation}
Indeed, this stopping criterion guarantees that the floor value of the
true error, defined in \S 3.1, is always reached.

   \begin{figure}
   \centering
   \includegraphics[width=10cm,angle=0]{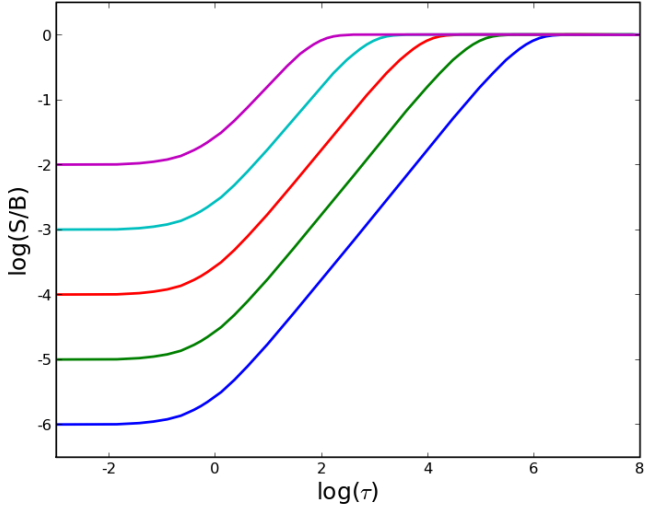}
   \caption{Solutions for the monochromatic scattering in an
     isothermal atmosphere, obtained for different values of
     $\varepsilon$, ranging from $10^{-4}$ to $10^{-12}$. Both the
     expected $\sqrt{\varepsilon}$ surface law and thermalisation
     lengths $\approx 1/\sqrt{\varepsilon}$ are perfectly recovered.}
         \label{Fig2}
   \end{figure}


\section{Results}

The solution of the non--LTE radiative transfer problem for a constant
property, plane-parallel 1D slab is a standard test for any new
numerical method. Indeed, the so-called \emph{monochromatic
  scattering} case allows a direct comparison of the numerical
solutions to the analytical ``solution'', $S_{\rm Edd}$, which can be
derived in Eddington's approximation (see e.g., the discussion in
\S5. of Chevallier et al. 2003).

\subsection{Scaling laws for non-LTE source function solutions}

In Fig.\,\ref{Fig2}, we display the run of the source function with
optical depth, for a \emph{self-emitting} slab of total optical
thickness $\tau_{\rm max}=2 \times 10^{8}$, using a 9-point per decade
spatial grid, a one-point angular quadrature with $\mu=\pm
1/\sqrt{3}$, and $\varepsilon$ values ranging from $10^{-4}$ to
$10^{-12}$.

Both the $\sqrt{\varepsilon}$ surface law and thermalisation lengths
$\approx 1/\sqrt{\varepsilon}$ are perfectly recovered, even for the
numerically difficult cases of large ($a \lessapprox 1$) albedos.

The ``true'' error, that is:

\begin{equation} {T}_{e} = {{\rm max}} \left( { {\mid S^{(j)} -
        S_{{\rm Edd}} \mid} \over {S_{\rm Edd}} } \right) \, ,
\end{equation}
is displayed in Fig.\,3, respectively for the GS, SOR and BiCG-P
iterative schemes. SOR was used with a $\omega=1.5$ over-relaxation
parameter, and $\varepsilon=10^{-12}$ in that case.

   \begin{figure}
   \centering
   \includegraphics[width=10cm,angle=0]{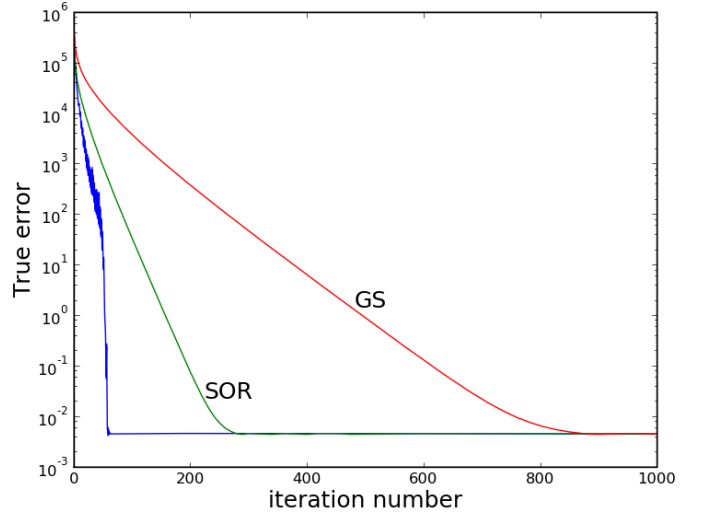}
   \caption{History of the true error vs. iteration number for
     $\varepsilon=10^{-12}$, respectively with Gauss-Seidel, SOR (with
     $\omega=1.5$) and BiCG-P with the diagonal of A as a
     preconditioner.}
         \label{Fig3}
   \end{figure}

\subsection{Timing properties}

The typical timing properties of the BiCG-P scheme, for the cases
considered here, are summarized in Table 1. While each GS/SOR
iteration is about 20\% longer than one ALI-iteration, the increase of
time per iteration, with the number of depth points, for BiCG-P is
significant. However, the balance between computing time and reduction
of the number of iterations always remains in favour of the BiCG-P
scheme vs. GS/SOR and, a fortiori, ALI.

\begin{table}
  \caption{Timing properties of the BiCG-P iterative schemes, depending on the optical depth grid refinement expressed as the number of point per decade, N$_{\tau}$, for a $\tau_{\rm max}=2 \times 10^{8}$, and
    $\varepsilon=10^{-8}$ self-emitting slab. We use a time normalized to the time for a ALI-iteration. Note that GS/SOR is, in comparison, about 20\% slower than ALI.}
  \centering
\begin{tabular}{c c c c c c c}
\hline\hline
  N$_{\tau}$ & 5 & 8 & 11 & 14 & 17 & 20 \\
\hline
  t/t$_{\rm ALI}$ & 0.85 & 1.2 & 1.8 & 2.1 & 2.3 & 2.7 \\
\hline
\end{tabular}
\end{table}

\subsection{Sensitivity to the spatial refinement}

We adopted the same slab model as the one reported in the previous
section, with $\varepsilon=10^{-8}$, and we allowed the grid
refinement to vary from 5 to 20 points per decade, with a step of 3.

   \begin{figure}
   \centering
   \includegraphics[width=10cm,angle=0]{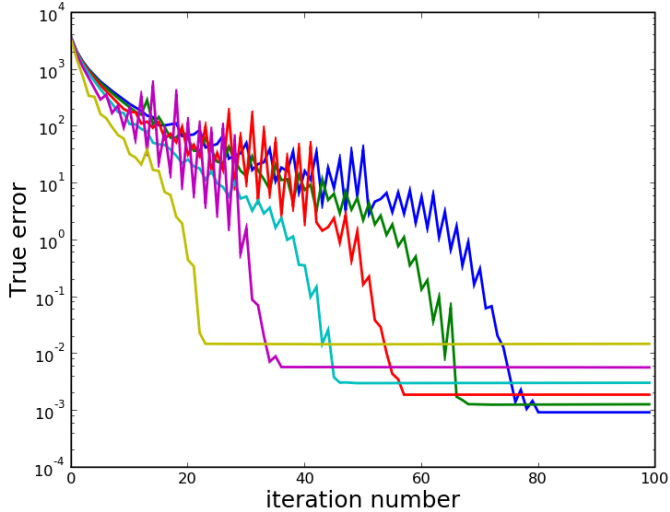}
   \caption{Evolution of the true error for varying spatial grid
     refinements, from 5 to 20 points per decade, with a step of
     3. The value of the true error \emph{plateau} decreases with an
     increasing grid refinement.}
         \label{Fig4}
   \end{figure}

   Figure\,4 shows the sensitivity of the method with grid
   refinement. Increasing grid refinement corresponds to a decreasing
   value of the floor value of the true error, which demonstrates the
   need of a sufficient spatial resolution for the sake of accuracy on
   the numerical solutions. The number of iterative steps necessary to
   fulfill the stopping criterion defined in Eq. (\ref{eq:stop_crit})
   is practically linear in the number of depth points.

\subsection{Smoothing capability}

A strong argument in favour of the Gauss-Seidel iterative scheme
resides in its smoothing capability. This point is particularly
important in the frame of multi-grid methods for the solution of
non--LTE radiative transfer problems (see e.g., Fabiani Bendicho et
al. 1997).

   \begin{figure}
   \centering
   \includegraphics[width=10cm,angle=0]{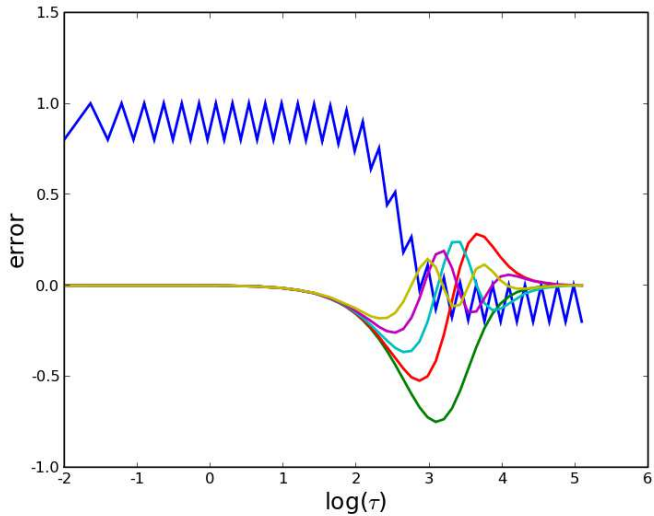}
   \caption{Initialization and evolution of the error,
     $S^{(j)}-S^{(\infty)}$, during the 5 first iterations of the
     BiCG-P scheme. The high-frequency component introduced \emph{ab
       initio} in the source function is removed after the very first
     iteration of BiCG-P. Other curves, of continuously decreasing
     amplitudes, correspond to iterations 2 to 5.}
         \label{Fig5}
   \end{figure}

   We have reproduced the test initially proposed by Trujillo Bueno \&
   Fabiani Bendicho (1995, see their Fig. 9) for the GS method. Here
   we consider a semi-infinite slab of maximum optical thickness
   $10^{5}$, sampled with a 9-point per decade grid, and
   $\varepsilon=10^{-6}$. The test consists in injecting \emph{ab
     initio} a high-frequency component into the error,
   $S^{(0)}-S^{(\infty)}$, on the initial guess for the source
   function. In Fig.\,5, the initial error we used is easily
   recognized as the hyperbolic tangent-like curve, to which a
   high-frequency component was added.

   After the very first iterate of the BiCG-P iterative process, the
   error has an amplitude of $\approx 0.75$ and, more importantly, it
   is already \emph{completely} smoothed. Other error curves displayed
   in Fig.\,5, of continuously decreasing amplitudes, are the
   successive errors at iterates 2 to 5.

   This simple numerical experiment demonstrates that BiCG-P can also
   be used as an efficient smoother for multi-grid methods.


\section{Conclusion}

We propose an alternative method for the solution of the non--LTE
radiation transfer problem. Preliminary but standards tests for the
two-level atom case in 1D plane parallel geometry are successful. In
such a case, the timing properties of the \emph{preconditioned} BiCG
method are comparable to the ones of ALI, GS and SOR stationnary
iterative methods. However the convergence rate of BiCG-P is
signifiantly better.

The main potential drawback of the BiCG-P method is in the usage of
the $A^{T}$ operator. A detailed trade-off analysis between such
over-computing time and an improved convergence rate, with respect to
the ones of GS/SOR iterative schemes, is currently being
conducted.

Multi-level and multi-dimensional cases will be considered
further. However, the use of BiCG-P does not require the cumbersome
modifications of multi-dimensional formal solvers required by
GS/SOR methods (see e.g., L\'eger et al. 2007).

\begin{acknowledgements}
  We are grateful to Drs. Bernard Rutily, and Lo\"{\i}c Chevallier
  for numerous discussions about the solution of the radiative
  transfer equation.
\end{acknowledgements}


\begin{thebibliography}{}

   \bibitem[2005]{amosov} Amosov, A.A. \& Dmitriev, V.V. 2005, MPEI
     Bulletin, 6, 5 (in russian)

   \bibitem[1969]{cl} Auer, L.H. \& Mihalas, D. 1969, ApJ, 158, 641

   \bibitem[1994]{lhafp} Auer, L.H. \& Paletou, F. 1994, A\&A, 285, 675

   \bibitem[1994]{lhafbtb} Auer, L.H., Fabiani Bendicho, P. \& Trujillo
     Bueno, J. 1994, A\&A, 292, 599

    \bibitem[1973]{cannon} Cannon, C.J. 1973, ApJ, 185, 621 

   \bibitem[2003]{loicetal} Chevallier, L., Paletou, F. \& Rutily, B. 2003,
      A\&A, 411, 221

    \bibitem[1967]{cuny} Cuny, Y. 1967, Ann. Ap., 30, 143 

    \bibitem[1997]{fabiani} Fabiani Bendicho, P., Trujillo Bueno,
      J. \& Auer, L.H. 1997, A\&A, 324, 161

    \bibitem[1996]{faurob} Faurobert, M., Frisch, H. \& Nagendra,
      K.N. 1997, A\&A, 322, 896

    \bibitem[1964]{feautrier} Feautrier, P. 1964, C.R.A.S., 258, 3189

    \bibitem[1952]{cg} Hestenes, M.R. \& Stiefel, E. 1952, Journal of
      Research of the National Bureau of Standards, 49(6), 409

   \bibitem[2007]{leger} L\'eger, L., Chevallier, L. \& Paletou, F. 2007,
      A\&A, 470, 1

    \bibitem[1978]{mihalas} Mihalas, D. 1978, Stellar Atmospheres (San
      Francisco: Freeman)

    \bibitem[1986]{oab} Olson, G.L., Auer, L.H. \& Buchler,
      J.R.. 1986, JQSRT, 35, 431


   \bibitem[1991]{mali1} Rybicki, G.B. \& Hummer, D.G. 1991,
      A\&A, 245, 171


   \bibitem[1981]{scharmer} Scharmer, G.B. 1981, ApJ, 249, 720

    \bibitem[2008]{saad} Saad, Y. 2008, Iterative methods for sparse linear
      systems (Philadelphia: SIAM)

    \bibitem[1995]{tf1} Trujillo Bueno, J., \& Fabiani Bendicho,
      P. 1995, ApJ, 455, 646

    \bibitem[1999]{trafa} Trujillo Bueno, J., \& Manso Sainz,
      R. 1999, ApJ, 516, 436

    \bibitem[2002]{noort} van Noort, M., Hubeny, I. \& Lanz, T. 2002,
      ApJ, 568, 1066
 
\end{thebibliography}
\end{document}